\documentclass{emulateapj}

\shorttitle{Molecular Outflows in the Substellar Domain}
\shortauthors{Phan-Bao et al.}

\begin{document}

\title{Molecular Outflows in the Substellar Domain: Millimeter Observations
of Young Very Low Mass Objects in Taurus and $\rho$ Ophiuchi}

\author{
Ngoc Phan-Bao,\altaffilmark{1,2}
Chin-Fei Lee,\altaffilmark{2}
Paul T.P. Ho,\altaffilmark{2,3}
Ya-Wen Tang\altaffilmark{2}}
\altaffiltext{1}{Department of Physics, HCMIU, Vietnam National University Administrative Building, Block 6, Linh Trung Ward, Thu Duc
District, HCM, Vietnam; pbngoc@hcmiu.edu.vn}
\altaffiltext{2}{Institute of Astronomy and Astrophysics, Academia Sinica, P.O. Box 23-141, Taipei 106, Taiwan, ROC; pbngoc@asiaa.sinica.edu.tw}
\altaffiltext{3}{Harvard-Smithsonian Center for Astrophysics, Cambridge, MA}

\begin{abstract}
We report here our search for molecular outflows from young very low mass stars and brown
dwarfs in Taurus and $\rho$ Ophiuchi. 
Using the Submillimeter Array (SMA) and the Combined Array for Research
in Millimeter-wave Astronomy (CARMA), we have observed 4 targets
at 1.3 mm wavelength (230 GHz) to search for CO J=2$\rightarrow$1 outflows.
A young very low mass star MHO 5 (in Taurus) with an estimated mass of 90 M$_{J}$,
which is just above the hydrogen-burning limit, shows
two gas lobes that are likely outflows. 
While the CO map of MHO 5 does not show a clear structure of outflow,
possibly due to environment gas,  
its position-velocity diagram indicates two distinct blue- and redshifted components.
We therefore conclude that they are components of a bipolar molecular
outflow from MHO 5. 
We estimate an outflow mass of $7.0 \times 10^{-5} M_{\odot}$ and a mass-loss rate of
$9.0 \times 10^{-10} M_{\odot}$. These values are over two orders of magnitude
smaller than the typical ones for T Tauri stars and somewhat weaker than
those we have observed in the young brown dwarf ISO-Oph 102 of 60 M$_{J}$ in $\rho$
Ophiuchi. This makes MHO~5 the first young very low mass star showing a bipolar molecular
outflow in Taurus. The detection boosts the scenario that very low mass objects
form like low-mass stars but in a version scaled down by a factor of over 100. 
\end{abstract}

\keywords{ISM: jets and outflows --- ISM: individual ([BHS98] MHO 5, [GY92] 3,
ISO-Oph 32, 2MASS J04390396+2544264, 2MASS J04414825+2534304) --- 
stars: formation --- stars: low mass, brown dwarfs --- technique: interferometric}

\section{Introduction}
Stars with a few solar masses 
can form by direct gravitational collapse mechanism (e.g., \citealt{shu87}).
The typical process of star formation 
starts with collapse, accretion and launching of material
as a bipolar outflow \citep{lada}. 
For the case of very low mass objects at the bottom of the main sequence, 
brown dwarfs (BD) (15--75~$M_{\rm J}$)
and very low-mass (VLM) stars (0.1--0.2 $M_{\odot}$) have
masses significantly below the typical Jeans mass ($\sim$1 M$_{\odot}$)
in molecular clouds, and hence it is difficult
to make a VLM stellar embyro by direct gravitational collapse 
but prevent subsequent accretion of material onto the central object once
the VLM embryo formed.
These VLM objects are therefore thought to form by different mechanisms 
(see \citealt{whitworth} and references therein).
Two major models have been proposed for their formation. 
In the standard formation model, they form like low-mass stars
just in a scaled-down version,
through gravitational collapse and turbulent fragmentation
of low-mass cores (e.g., \citealt{padoan}).
In the ejection scenario, the VLM objects are simply 
stellar embryos ejected from unstable multiple 
protostellar systems by dynamical interaction
with the other embyros.
These VLM embryos are ejected from their gas resevoir 
and then they become VLM stars and BDs
(e.g., \citealt{reipurth,bate02,bate05}).

Observations (see \citealt{luhman07} and references therein) 
of the BD and VLM star properties in different star-forming
regions such as their initial mass function, 
velocity and spatial distributions, multiplicity, accretion disks and jets
have demonstrated that stars and BDs share similar properties.
This strongly supports the scenario that BDs and VLM stars form 
as low-mass stars do. One should note that additional mechanisms
(e.g., the ejection) are possible
but they are not likely dominant in making VLM objects.

More observations are needed to understand how these VLM objects form, especially
observations at very early stages provide us an insight into the VLM object
formation mechanism.
\citet{pb08} reported the first detection of a 
bipolar molecular outflow from the young BD ISO-Oph~102 ($\sim$60~$M_{\rm J}$)
in the $\rho$ Ophiuchi dark cloud. The
outflow mass and the mass loss rate are over 2 orders of magnitude
smaller than the typical values detected for T Tauri stars. Their result demonstrates
that the molecular outflow process occurs in BDs
as a scaled-down version of that seen in low-mass stars.
As the molecular outflow is a fundamental part of the star formation process, 
the detection of molecular outflows in VLM stars and BDs therefore
extends our understanding of star formation 
in the lowest mass domain.

Up to date, there are only two molecular outflows detected in the substellar domain: 
from L1014-IRS, a possible proto-BD \citep{bourke,huard} and ISO-Oph 102, a young BD \citep{pb08}.
In this paper, we present our millimeter observations of
3 BDs and 1 VLM star in Taurus and $\rho$ Ophiuchi.
We detected a bipolar molecular outflow from the young VLM star MHO~5
whose outflow properties are similar to those we observed in ISO-Oph 102. 

Sec.~2 presents our sample,
Sec.~3 reports our millimeter observations and the data reduction,
Sec.~4 discusses outflow properties observed in MHO 5 and Sec.~5 summarizes our results.

\section{Sample selection}
Following the discovery of a bipolar molecular
outflow in ISO-Oph 102, we selected 4 additional 
targets in $\rho$ Ophiuchi and Taurus (see Table~\ref{log}).
They were selected from the lists of VLM accretors
given in \citet{muz03,muz05} for Taurus and \citet{natta04} for
$\rho$ Ophiuchi to cover a wide range of mass
from 35 to 90 M$_{\rm J}$, one of them (MHO 5) is a very
low-mass star. These sources are the strongest accretors for a given mass. 
In addition, the detection of signatures of outflows from the literature
was also considered to select the sources. 
Two of them MHO 5 and ISO-Oph 32 show forbidden emission lines (FELs)
that could be associated with outflow activities. 
The detection of the blueshifted jet component 
of ISO-Oph 32 has been reported in \cite{whelan09}.
They are therefore good
targets for our molecular outflow search. 

\section{Observations and data reduction}
We observed 4 VLM objects in Taurus and $\rho$ Ophiuchi 
with SMA and CARMA.
The observing log is given in Table~\ref{log}.

\subsection{SMA observations}
Observations of MHO~5 and ISO-Oph 32 were carried out
with the receiver band at 230~GHz of the SMA\footnote{
The Submillimeter Array is a joint project between the 
Smithsonian Astrophysical Observatory and 
the Academia Sinica Institute of Astronomy and Astrophysics 
and is funded by the Smithsonian Institution and the Academia Sinica.} \citep{ho}
on November 16$^{\rm th}$ 2008 and June 8$^{\rm th}$ 2009, respectively.
Zenith opacities at 225 GHz were typically in the range 
0.15--0.2 and 0.05--0.1 for MHO 5 and ISO-Oph 32, respectively.
Both 2 GHz-wide sidebands, which are separated
by 10 GHz, were used. The SMA correlator was configured with high spectral
resolution bands of 512 channels per chunk of 104~MHz for $^{12}$CO, $^{13}$CO, 
and C$^{18}$O $J=2 \rightarrow 1$ lines, giving a channel spacing of 0.27~km~s$^{-1}$.
A lower resolution of 3.25~MHz per channel was set up for the remainder of
each sideband. The quasars 0428+329, 0510+180 and 3C~454.3 
have been observed for gain and passband calibration
of MHO~5, respectively. 
In the case of ISO-Oph 32, the quasars 1625$-$254 and 3C~273
were used. Uranus was used for flux calibration for both the targets. 
The uncertainty in the absolute flux calibration is
about 10\%. 
The data were calibrated using the MIR software package
and further analysis was carried out with the MIRIAD package adapted
for the SMA. 
All 8 antennas were operated in the compact configuration, 
resulting in synthesized beams of 
3$''$.05~$\times$~2$''$.82 and 3$''$.78~$\times$~3$''$.38
using natural weighting for
MHO~5 and ISO-Oph 32, respectively. 
The primary FWHM  
beam is about 50$''$~at the observed frequencies.
The rms sensitivity was about 1~mJy for the continuum,
using both sidebands and $\sim$0.2~Jy~beam$^{-1}$ per channel for
the line data. 
We did not detect the dust continnum emission with an upper limit of 3~mJy
measured at the VLM star position.

\subsection{CARMA observations}
The two least massive BDs 2M~0439 and 2M~0441 in our sample
were observed with CARMA in February and March 2009.
All six 10.4-meter, nine 6.1-meter antennas were operated
in the D configuration (a beam size of 2.4$''$ $\times$ 2.0$''$) 
at 230 GHz. 
Zenith opacities at 227 GHz were typically in the range 
0.3--0.4 and 0.2--0.3 for 2M~0439 and 2M~0441, respectively.
As CARMA with a larger collecting area 
is more sensitive than SMA, the array is therefore suitable for
our outflow search in these BDs.
In semester 2009A, 
all three 500 MHz-wide bands (1.5 GHz maximum bandwidth per sideband), 
which may be positioned independently with the IF bandwith, 
were used for CO search with different spectral resolutions.
For our observations, three bands were configured in the following modes: 
8 MHz and 31 MHz with 63 channels per band, 
and 500 MHz with 15 channels per band,
resulting in spectral resolutions of 0.17, 0.64, 
and  43.4 km~s$^{-1}$, respectively.
The quasar 3C 111 has been observed for gain, 
and 3C 84 for passband and flux calibration.
The uncertainty in the absolute flux calibration is
about 10\%.
The data were reduced with the MIRIAD package adapted for the CARMA. 
The synthesized beam sizes are about 2$''$.60~$\times$~2$''$.48 
and 2$''$.02~$\times$~1$''$.86 (natural weighting)
for 2M~0439 and 2M~0441, respectively.
The primary FWHM beam is about 36$''$ at 230 GHz.
For the continuum, the rms sensitivity was about 0.4~mJy.
For the line data, the sensitivities are 
0.1 and 0.05~Jy~beam$^{-1}$ per channel
for 8 and 31 MHz band, respectively.
\section{Discussion}

\subsection{2M 0439 and 2M 0441}
Two young BDs in Taurus 2MASS J04390396+2544264 (M7.25, 50 M$_{\rm J}$, \citealt{lu04,muz05})
and 2MASS J04414825+2534304 (M7.75, 35 M$_{\rm J}$, \citealt{lu04,muz05}) 
are strong accretors with an estimated accretion rate 
$\log\dot{M_{\rm acc}}(M_{\odot}$~yr$^{-1})=-11.3$ for both the BDs 
\citep{muz05}.
These BDs are included in our sample with an aim of probing molecular outflows
for the lowest mass BDs. Up to date, 
no outflow signatures from these two BDs have been reported.  

The dust continuum fluxes are 2.4$\pm$0.4 mJy and 2.2$\pm$0.4 mJy measured 
at the position of 2M 0439 and 2M 0441, respectively. Our measurements
are consistent with the previous ones \citep{scholz06}.

Our CO maps from CARMA data do not reveal any outflow. 
The non-detection of outflows indicates three possibilities: 
(1) the outflow process in these BDs itself is too weak to produce
detectable outflows; (2) this process has already stopped; (3) there is not
enough environment gas around the sources to produce detectable outflows.
\citet{guieu} reported the presence of an inner hole in these BD disks 
(see table 4), indicating these BDs in the transitional stage between
class II and III objects, which are at later stages
than that of embedded proto-BD candidates such as L1014-IRS (class I, \citealt{bourke})
or SSTB213 J041757 (class I, \citealt{barr09}). 
This therefore supports the second scenario. 
However one should note that 
the outflow process in BDs might last longer than that in low-mass stars
as seen in ISO-Oph~102 \citep{pb08}. Therefore the first 
scenario can not be ruled out. 

\subsection{ISO-Oph 32}
ISO-Oph 32 (or [GY9] 3) 
is an M7.5 accreting BD (class II) with an estimated mass of 40~M$_{\rm J}$ 
and an accretion rate $\log\dot{M}_{\rm acc}(M_{\odot}$~yr$^{-1})=-10.5$
estimated from H$\alpha$ profiles \citep{natta04}.
The dust continuum is not detected and we obtain an upper limit
of 1.0~mJy at the BD position.

\citet{whelan09} reported the detection of a weak jet in this BD from the 
[O{\small I}] 6300\AA~line at a signal-to-noise ratio of just above 4.
However, the CO map of ISO-Oph~32 based on our SMA observations
does not show any outflow, possibly due to either the molecular outflow process
is too weak to be detectable with our current sensitivity
or surrounding gas is not dense enough to produce detectable molecular 
outflows. 
Deeper observations
are required to confirm the previous detection of a jet and to characterize the molecular
outflow from this BD.

\subsection{MHO 5}
MHO~5, a young ($\sim$1--2 Myr) M6 dwarf
initially identified by \citet{briceno}, is 
located in the Taurus dark cloud at a distance of 
147~parsecs \citep{loi}.  
\citet{briceno} (see also \citealt{muz03}) detected 
forbidden [O{\small I}] emission lines at 6300
and 6364~\AA. Such forbidden emission lines are typically associated
with accretion disks or outflows \citep{alencar,hartigan}.
The estimated mass is 90~$M_{\rm J}$ \citep{muz03}, which is 
just above the hydrogen-burning limit.
The optical visibility of MHO~5 suggests that the source 
corresponds to a class II object \citep{lada} in the star formation phase, 
a class with an accreting circumstellar disk and the protostar 
at this stage is the so-called 
classical T Tauri star.

\begin{figure}
\vskip 1in
\hskip -0.25in
\centerline{\includegraphics[width=5in,angle=-90]{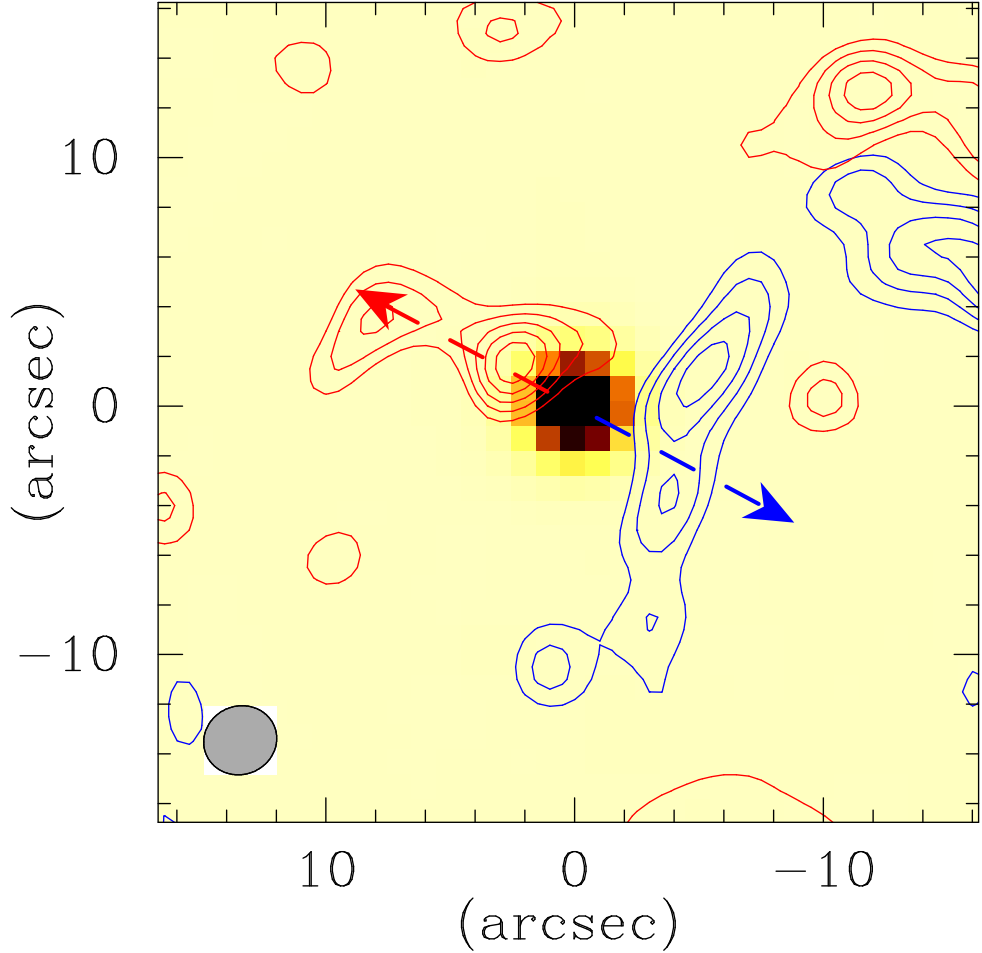}}
\caption{\normalsize An overlay of the J-band (1.25 $\mu$m) near-infrared 
Two Micron All Sky Survey (2MASS) image and
the integrated intensity in the carbon monoxide (CO $J=2-1$) line 
emission from 2.45 to 6.15~km~s$^{-1}$ line-of-sight velocities.
The blue and red contours represent the blueshifted 
(integrated over 2.45 and 3.77~km~s$^{-1}$) and redshifted 
(integrated over 4.56 and 6.15~km~s$^{-1}$)
emissions, respectively. 
The contours are 4, 6, 8,...times the 
rms of 0.12~Jy~beam$^{-1}$~km~s$^{-1}$. The brown dwarf is visible in the
J-band image. The position angle of the outflow is about 62$^{\circ}$. 
The outflow directions are indicated by the blue and red arrows.
The blue- and redshifted gas lobes are displaced on opposite
sides of the brown dwarf center with an offset of about 2$''$.
An elongated structure is visibly seen in the blueshifted component
while the redshifted one shows an extended component (see Sec.~4.3 for
discussion). 
The gas lobes in the top right corner are expected to be outflows
from neaby massive outflows in the L1551 molecular cloud
(see Fig. 4 and Fig. 15 in \citealt{mori}).
The synthesized beam is shown in the bottom
left corner.
\label{map}}
\end{figure}

\begin{figure}
\vskip 1in
\hskip -0.25in
\centerline{\includegraphics[width=4in,angle=-90]{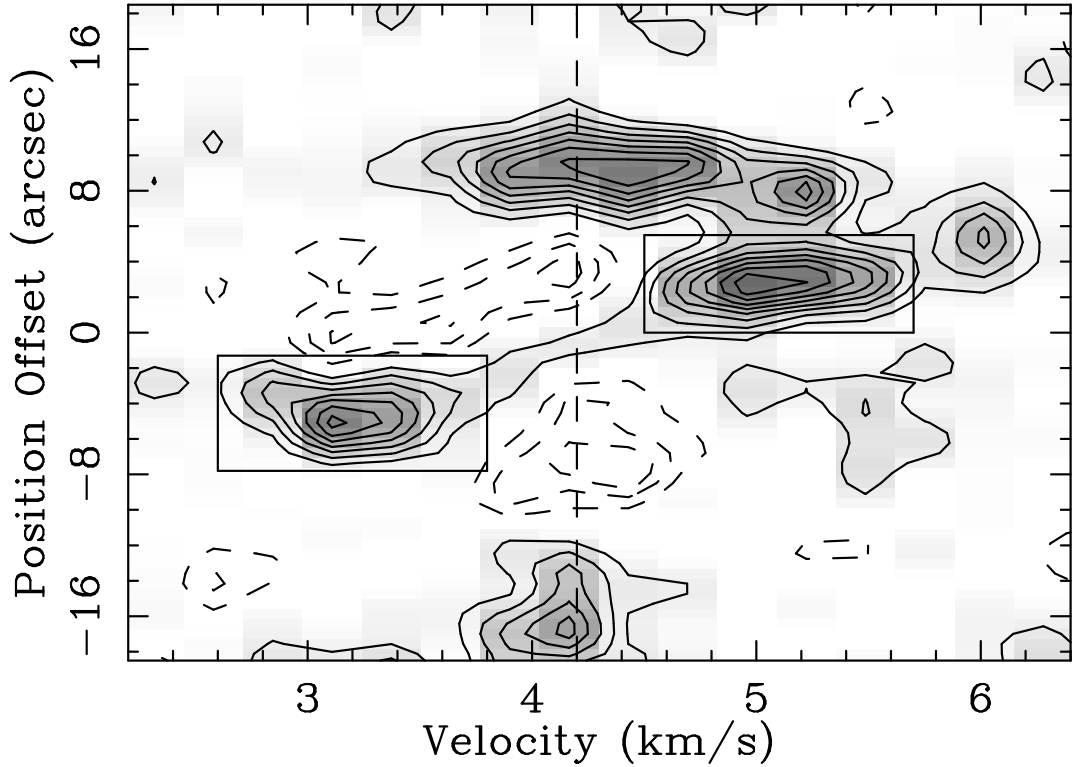}}
\caption{\normalsize Position-Velocity (PV) cut diagram for CO~$J=2 \rightarrow 1$ emission
at a position angle of 62$^{\circ}$.
The contours are $-$8, $-$6, $-$4, $-$2, 2, 4, 6, 8,...times the 
rms of 0.1 Jy beam$^{-1}$. 
The blueshifted (2.6-3.8~km~s$^{-1}$) and redshifted (4.5-5.7~km~s$^{-1}$)
components expected from
MHO 5 are indicated by boxes, which we take to estimate the MHO 5
outflow properties (see Sec.~4.3 for discussion).
The gas velocity at the source position
is 4.2$\pm$0.3~km~s$^{-1}$, which we take to be the systemic velocity 
of the VLM star
as indicated by the dashed line. 
Both blue- and redshifted components shows a wide range of
the velocity in their structure, which appears 
to be the bow-shock surfaces as observed in young stars \citep{lee}. These
surfaces are formed at the head of the jet and accelerate the material
in the bow-shock sideways (e.g., \citealt{masson}).
\label{pv}}
\end{figure}

Figure~\ref{map} presents an overlay of a near-infrared image and 
the integrated intensity in the carbon monoxide (CO $J=2-1$) line 
emission. Two blue- and redshifted
CO lobes are detected around the position of the central
object. Both lobes show elongated structures
and but they are not symmetrically displaced
on opposite sides of the MHO~5 position
as was seen in ISO-Oph~102.
However the position-velocity (P-V) diagram for the CO emission
cut at a position angle of
62$\degr$ clearly indicates two blue- and redshifted components
(Fig.~\ref{pv}).
The elongated structures are probably due to a mixture of
outflows from MHO~5 and nearby massive outflows from class 0 and I protostars
in the L1551 molecular cloud, a multigenerational star formation
region (see \citealt{mori}).
Based on the P-V diagram and assuming that the MHO 5 outflow
components are symmetric, we consider the outflow
from MHO~5 to consist of only the blueshifted component with the peak
at velocity $\sim$3 km~s$^{-1}$ and the redshifted component with the peak
at velocity $\sim$5 km~s$^{-1}$ as indicated Fig.~\ref{pv}.
The CO gas velocity at the source position is 4.2 km~s$^{-1}$
and our spectral resolution is 0.27 km~s$^{-1}$, we therefore estimate
the systemic velocity of MHO 5 to be about 4.2$\pm$0.3 km~s$^{-1}$.
This value is in good agreement with a velocity
of 4.0 km~s$^{-1}$, which is determined by taking an average 
of the velocities of red- and blueshifted components
as used to determine the systemic velocity of ISO-Oph 102
\citep{pb08}. One should note that these estimates 
suffer some uncertainties due to some problems such as
missing flux, outflow contamination
for the first estimate or outflow morphology for the second one.
More observations are needed to exactly determine
the systemic velocity.  
If this systemic velocity is shifted by 
the uncertainty of 0.3 km~s$^{-1}$ in either direction, 
the blueshifted and redshifted contours will not significantly change as 
the MHO 5 outflow shows two distinct outflow components 
in the velocity ranges 2.6--3.8 km~s$^{-1}$
and 4.5--5.7 km~s$^{-1}$, which are not affected by the shift
of systemic velocity.
One should note that the origin of extended redshifted components
around 4.4 and 5.2 km~s$^{-1}$ at an offset position
of $\sim$8$''$ and around 6.0 km~s$^{-1}$ at a position
of $\sim$6$''$
(see Fig.~\ref{pv}) is not clear, we therefore do not
include them in our outflow mass calculation.
However, if these extended components are included in the
MHO~5 outflow, its outflow mass will increase by a factor of $\sim$2.
This increase in outflow mass however will not change our main
conclusion on the outflow properties of MHO~5 comparing to those
of low-mass stars.

We then measure the size of each lobe of about 4$''$ corresponding
to $\sim$600~AU in length. 
The two outflow components likely show a bow-shock structure
as seen in ISO-Oph 102, an effect of the interaction
between the jet propagation and the ambient material
(e.g., \citealt{lee,masson}). 
Following the standard manner \citep{cabrit,andre} as used in the
previous paper \citep{pb08}, we calculate the
outflow properties. We use a value of 35~K \citep{loren}
for the excitation temperature, and we derive a lower limit to 
the outflow mass $M_{\rm flow} \sim 1.4 \times 10^{-5} M_{\odot}$. 
If we correct for optical depth
with a typical value of 5 \citep{l88a}, we obtain an upper limit
to the outflow mass of $7.0 \times 10^{-5} M_{\odot}$. 
One should note that a missing flux factor for SMA
is not applied for the upper limit.
This value is smaller than the typical values 0.01--0.7 $M_{\odot}$ 
of young low-mass stars (class II with G, K spectral types,
see \citealt{l88b} and references therein) by over three orders of magnitude.

We use the observed maximum flow velocity of 1.8 km~s$^{-1}$
and apply a correction for an outflow inclination, which is
derived from a disk inclination of $\sim$30$\degr$ (B. Riaz, private communication),  
to compute upper limit
values for the kinematic and 
dynamic parameters. 
We find that the momentum is $P=2.8 \times 10^{-5}$$M_{\odot}$~km~s$^{-1}$,
the energy is $E=3.0 \times 10^{-5}$$M_{\odot}$~km$^{2}$~s$^{-2}$, 
the force is $F=1.8 \times 10^{-8}$$M_{\odot}$~km~s$^{-1}$~yr$^{-1}$,
and the mechanical luminosity is $L=3.1 \times 10^{-6}$~L$_{\odot}$, where L$_{\odot}$
is the solar luminosity. A correction for the optical depth factor of 5 and 
a missing flux factor of 3 for SMA \citep{bourke} will increase these upper limit values 
by a factor of 15. Using the lower value of the outflow mass and
a correction factor of 10 applied for the outflow duration time \citep{parker},
we derive the outflow mass-loss rate 
$\dot{M}_{\rm out}=9.0 \times 10^{-10}$$M_{\odot}$~yr$^{-1}$, which is smaller
than a typical value of $10^{-7}$$M_{\odot}$~yr$^{-1}$
for T Tauri stars (see \citealt{lada}) by two orders of magnitude.
One should note that the correction factors were also
applied in the same manner to calculte the outflow properties of ISO-Oph~102
\citep{pb08}.
The outflow mass and the mass loss rate values from MHO~5 are
somewhat weaker than that from ISO-Oph 102 but they
are comparable to each other. 
This can be explained by a weaker activity of the jet in MHO~5 than in ISO-Oph 102,
e.g., [S{\small II}] 6716 and 6731~\AA~ only detected 
in ISO-Oph 102 \citep{briceno,natta04}, 
and a lower gas density surrounding MHO~5 (Taurus)
than in ISO-Oph~102 ($\rho$ Ophiuchi) (see \citealt{tachi}).
One should note that the outflow of these
sources share similar outflow mass and mass loss rate
with the proto-BD candidate L1014-IRS, 
$M_{\rm flow} \sim 1.4 \times 10^{-5}$~$M_{\odot}$ and 
$\dot{M}_{\rm out}=2 \times 10^{-9}$~$M_{\odot}$~yr$^{-1}$ \citep{bourke}.
The accretion and mass loss rates for these sources
are listed in Table~\ref{mloss}.

We also examine the possibility that the emission might be due to bound motions
and not outflow emission. This would require
an enclosed mass of 1.1~$M_{\odot}$ for an outflow size of
600~AU with a velocity of 1.8~km~s$^{-1}$, which is significantly larger
than the core mass of $< 0.2$~$M_{\odot}$ within the same radius \citep{onishi}.
We therefore conclude that the detected emission is from the outflow.

\begin{figure}
\vskip 1in
\hskip -0.25in
\centerline{\includegraphics[width=6in,angle=-90]{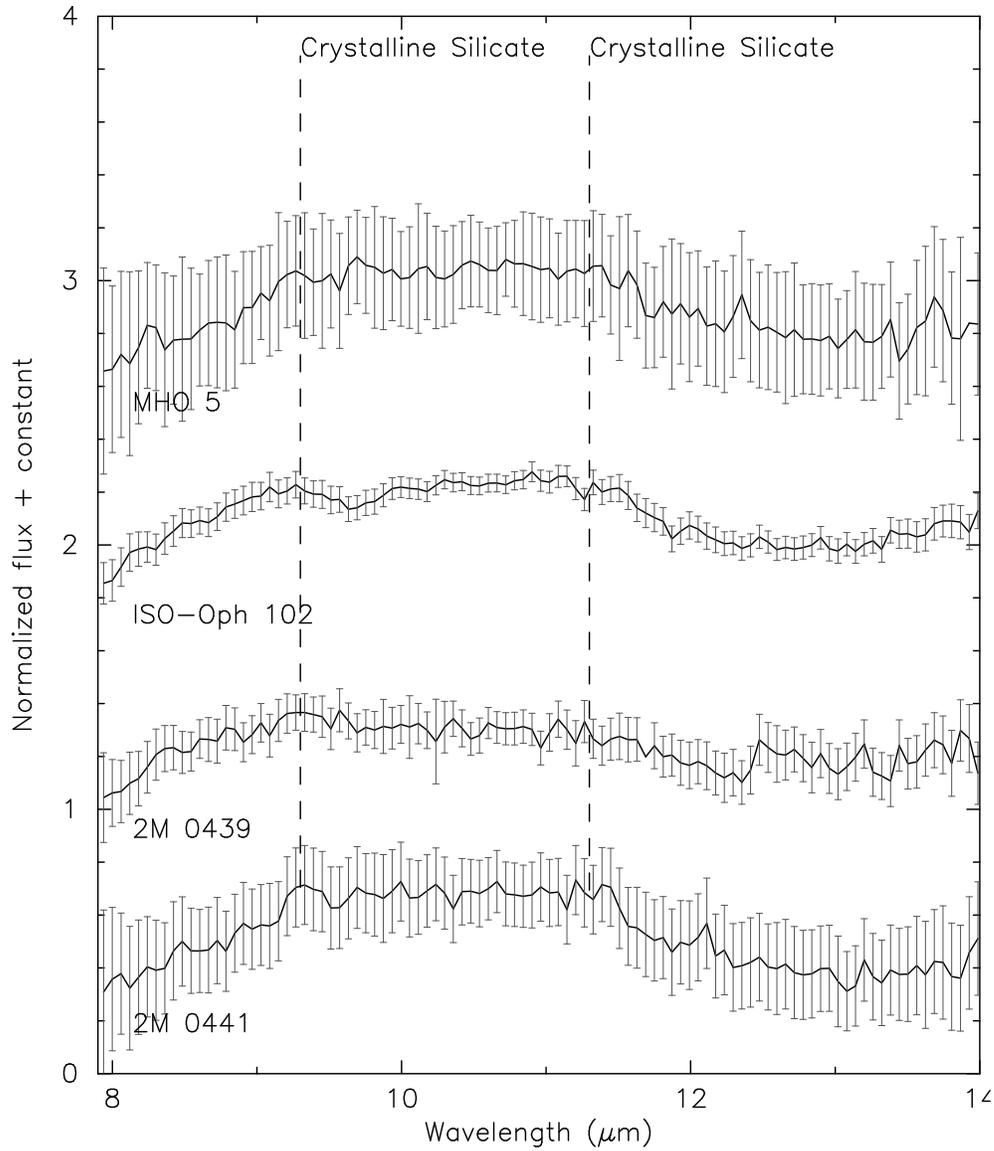}}
\caption{\normalsize Spitzer infrared spectra of MHO~5, ISO-Oph 102, 2M 0439 
and 2M 0441. Continuum subtraction was done following the literature 
\citep{van03,apai}. The crystalline silicate features at 9.3
mm (mainly enstatite) and 11.3 mm (forsterite) are indicated.
\label{crys}}
\end{figure}

We also reduced and analyzed infrared observations of MHO~5, 2M 0439, and 2M 0441 
from the Spitzer Space Telescope's archival data. 
The IRS infrared (SL: 5.2-14.5 $\mu$m; LL: 14.0-38.0 $\mu$m) \citep{houck}
spectra of the three sources, were produced from
the basic-calibrated data downloaded from the Spitzer archive 
(PID: 30540, 2)
and reduced using the SMART software package \citep{higdon}.
All their mid-infrared spectra (see Fig.~\ref{crys}) exhibit strong crystalline
silicate features (enstatite at 9.3~$\mu$m and forsterite at 11.3~$\mu$m),
which appear similar to that seen in ISO-Oph 102 
\citep{pb08} and other VLM stars and BDs \citep{apai,merin,riaz09}.
The strong exhibition of crystalline
silicates provides a direct evidence of grain growth and dust settling
in the VLM star and BD disk. 
In the case of MHO~5, the molecular outflow process, which sweeps out gas
from the disk, coexisting with grain growth and 
crystallization may favor rocky planets forming around
this VLM star. One should note that
this coexistence phenomenon 
has also been observed in ISO-Oph 102.

\section{Summary}
Here, we report our search for molecular outflows
from VLM stars, BDs 
and the discovery of the bipolar molecular outflow
in the young VLM star MHO~5. We point out that
the bipolar molecular outflow in VLM stars is very similar
to outflows as seen in young stars but scaled down by three and two 
orders of magnitude for the outflow mass and the mass-loss rate, respectively.
This additional evidence strongly supports the scenario 
that VLM stars, BDs, and perhaps young planetary mass objects can launch a  
bipolar molecular outflow. If so, these objects may have 
a common origin with the low-mass stars. 
The detection of the coexistence phenomenon of molecular outflow, 
grain growth and crystallization processes in MHO~5 and ISO-Oph 102
provides an important implication for rocky planet search around VLM stars
and BDs.

\acknowledgments
N.P.-B has been supported by VietNam NAFOSTED grant 103.08-2010.07.
Support for CARMA construction was derived from the Gordon and Betty Moore Foundation, the Kenneth T. and Eileen L. Norris Foundation, the James S. McDonnell Foundation, the Associates of the California Institute of Technology, the University of Chicago, the states of California, Illinois, and Maryland, and the National Science Foundation. Ongoing CARMA development and operations are supported by the National Science Foundation under a cooperative agreement, and by the CARMA partner universities. 
This work is based in part on observations made with the Spitzer 
Space Telescope, which is operated by the Jet Propulsion Laboratory,
California Institute of Technology, under a contract with NASA.
This work has made use of the Centre de Donn\'ees astronomiques de Strasbourg (CDS) 
database.

\clearpage

\clearpage

\clearpage

\clearpage

\begin{deluxetable}{llccl}
\tablewidth{0pt}
\tablecaption{Observing logs for the 4 young BDs and VLM stars 
  \label{log}}
\tablehead { 
\colhead {Target} & \colhead {Array} & \colhead {Configuration} & \colhead {Beam size} & 
\colhead {Region} \\
\colhead {}       & \colhead {} & \colhead {} & \colhead {($\arcsec \times \arcsec$)} & 
\colhead {} 
}
\startdata
 MHO 5       &  SMA   &  Compact & 3.05$\times$2.82 &  Taurus    \\
 2M 0439     &  CARMA &  D       & 2.60$\times$2.48 &  Taurus   \\
 2M 0441     &  CARMA &  D       & 2.02$\times$1.86 &  Taurus   \\
 ISO-Oph 32  &  SMA   &  Compact & 3.78$\times$3.38 &  $\rho$ Ophiuchi   \\
\enddata
\end{deluxetable}

\clearpage

\begin{deluxetable}{lcrll}
\tablewidth{0pt}
   \tablecaption{Accretion and mass loss rates from MHO 5 (Taurus) and ISO-Oph 102 
   ($\rho$ Ophiuchi), and the mass loss rate from 
    the embedded proto-BD candidate L1014-IRS \label{mloss}}
\tablehead {    
\colhead {Target} & \colhead {Mass ($M_{\rm J}$)} & \colhead {log($\dot{M}_{\rm acc}$)} & 
\colhead {log($\dot{M}_{\rm out}$)} &  \colhead {References$^{\rm a}$}  \\
}
\startdata 
ISO-Oph 102 & 60    &  $-$8.9    & $-$8.9   &   \citet{natta04,pb08}  \\
MHO 5       & 90    &  $-$10.8   & $-$9.1   &   \citet{muz03}; this paper  \\
L1014-IRS   &       &            & $-$8.7   &    \citet{bourke} \\
\enddata
\tablenotetext{a}{References for accretion and mass loss rates}
\end{deluxetable}

\end{document}